\documentclass[lettersize,journal]{IEEEtran}
\usepackage{amsmath,amsfonts}
\usepackage{algorithmic}
\usepackage{algorithm}
\usepackage{array}
\usepackage[caption=false,font=normalsize,labelfont=sf,textfont=sf]{subfig}
\usepackage{textcomp}
\usepackage{stfloats}
\usepackage{url}
\usepackage{verbatim}
\usepackage{graphicx}
\usepackage{cite}
\usepackage{bm}
\usepackage{color}
\newcommand{\ris}{\text{RIS}}
\newcommand{\bs}{\text{BS}}
\newcommand{\ue}{\text{UE}}
\newcommand{\br}{\text{B-R}}
\newcommand{\ru}{\text{R-U}}
\newcommand{\bfr}{\mathbf{r}}
\newcommand{\bfh}{\mathbf{h}}
\newcommand{\C}{\text{C}}
\newcommand{\cb}{}  

\begin{document}

\title{Near-Field Wideband Beamforming for RIS\\ Based on Fresnel Zone}

\author{Qiumo Yu, Linglong Dai,~\IEEEmembership{Fellow,~IEEE,}

	\thanks{All authors are with the Department of Electronic Engineering, Tsinghua University, Beijing 100084, China, and also with the Beijing National Research Center for Information Science and Technology (BNRist), Beijing 100084, China. (e-mails: yqm22@mails.tsinghua.edu.cn;
	daill@tsinghua.edu.cn).}

\thanks{This work was funded in part by the National Science Fund for Distinguished Young Scholars (Grant No. 62325106), and in part by the National Key R\&D Program of China (No. 2023YFB811503). 
}
}


\maketitle

\begin{abstract}
	Reconfigurable intelligent surface (RIS) has emerged as a promising solution to overcome the challenges of high path loss and easy signal blockage in  millimeter-wave (mmWave) and terahertz (THz) communication systems. With the increase of RIS aperture and system bandwidth, the near-field beam split effect emerges, which causes beams at different frequencies to focus on distinct physical locations, leading to a significant gain loss of beamforming. To address this problem, we leverage the property of Fresnel zone that the beam split disappears for RIS elements along a single Fresnel zone and propose beamforming design on the two dimensions of along and across the Fresnel zones. The phase shift of RIS elements along the same Fresnel zone are designed aligned, so that the signal reflected by these element can add up in-phase at the receiver regardless of the frequency. Then the expression of equivalent channel is simplified to the Fourier transform of reflective intensity across Fresnel zones modulated by the designed phase. Based on this relationship, we prove that the uniformly distributed in-band gain with aligned phase along the Fresnel zone leads to the upper bound of achievable rate. Finally, we design phase shifts of RIS to approach this upper bound by adopting the stationary phase method as well as the Gerchberg-Saxton (GS) algorithm. Simulation results validate the effectiveness of our proposed Fresnel zone-based method in mitigating the near-field beam split effect.
\end{abstract}


\section{Introduction}

With the emergence of new applications such as virtual reality, holographic images, and digital twins, communications have put forward requirements for high transmission rates \cite{chen_terahertz_2021}. Emerging technologies such as millimeter-wave (mmWave) and terahertz (THz) communications are expected to be adopted in the future, offering ultra-wide bandwidth of several gigahertz (GHz) or even higher to facilitate high-speed data transmission \cite{xiao_millimeter_2017,rappaport_wireless_2019}. However, mmWave and THz signals are easy to be blocked by objects in their propagation path, resulting in limited coverage range \cite{he_modeling_2023}. The recent development of reconfigurable intelligent surfaces (RIS) presents a solution to the blockage issue, as it can provide an additional reflection link by beamforming \cite{yang_blockage-aware_2024,zhang2023reconfigurable}. Meanwhile, high path loss poses a challenge for mmWave and THz techniques \cite{rappaport_wideband_2015,akyildiz_combating_2018}, necessitating high beamforming gain to mitigate the loss. In order to achieve high beamforming gain \cite{li_modular_2023,wei_codebook_2022},  RIS usually consists of extremely large number of elements (2304 elements \cite{cui_demo_2022}, for example).

\subsection{Prior Works}
As the scale of RIS continues to increase, the future RIS-enabled  wideband communication systems will work in near-field wideband scenario, facing issue on near-field beam split, which draws challenges for the system design, especially for RIS beamforming.  Specifically, the beam split effect was initially examined in far-field {\cb \cite{dai_reconfigurable_2020}}, resulting from the frequency-dependent nature of channels and the frequency-independent phase shift capability of RIS elements. This mismatch causes the beam direction to vary with frequency, leading to reduction in signal intensity at the user's location and degradation of performance. {\cb As the scale of RIS increases significantly, the far-field channel model is no longer accurate and should be replaced by near-field model where the  electromagnetic field is modeled as spherical wave \cite{an2024near-field, wang2024performance}}.  In near-field wideband systems, beam split exhibit significantly different characteristics than in the far-field. Beams of different frequencies in near-field wideband systems are focused on different physical areas in the 3-dimensional space \cite{cui_near-field_2021}, while the far-field beam split only in the angular domain. The near-field beam split effect causes the beams to deviate from the area where the user is located, seriously affecting the energy received by the user \cite{cui_near-field_2023}. Further, authors in  \cite{myers_infocus_2022} show that near-field beam split can affect the receiving power even in the bore sight direction where no far-field beam split effect occurs. 

Although many work on the issue of beam split have arose, only very few recent publishments investigated the near-field beam split of RIS. An intuitive solution is to adopt frequency-dependent modules, like {\cb true-time-delay (TTD)} and delay adjustable metasurface (DAM), to adjust the frequency-dependent channel.  A joint design of phase shifts and delay of the DAM was proposed in  \cite{hao_far-near-field_2023} to focus the beam of all frequency to the user's location, effectively eliminating the gain loss caused by beam split. {\cb Besides, an electromagnetically induced transparency (EIT) based adjustable-delay RIS is utilized to improve the wideband performance in orthogonal frequency division multiplexing (OFDM) communications \cite{an2024adjustable}.}  Further, to reduce the number of high-cost time delay unit, a sub-connected TTD architecture was proposed in  \cite{wang_wideband_2024}, where an end-to-end optimization for beamforming  is realized using neural network. 

While significant attention has been given to the issue of beam split, the specific phenomenon of near-field beam split in RIS has only been addressed in a few recent publications. First, the insertion loss of most mmWave TTD module design exceeds 10 dB \cite{govind_ultra-compact_2024,garakoui_compact_2015,jung_compact_2020}. This substantial loss results in a reduction of beamforming gain by over 10 times, significantly diminishing the benefits of TTD adoption. Secondly, configurable TTD modules typically incorporate dozens of electronic components such as switches, transmission lines, and couplers, among others. In contrast, an RIS element only comprises a PIN diode, a patch antenna and a small amount of control circuits, as discussed in \cite{yang_ris-aided_2021}. In the context of extremely-large-scale RIS implementations, the TTD structure entails an unbearable hardware cost in comparison to traditional designs. Thirdly, current TTD structures rely on the transmission line \cite{park_1540_2013}. The \textit{transmissive} structure is hard to adopted on the commonly-used \textit{reflective} RIS. As of now, literature only showcases non-configurable reflective array design that cannot freely adjust beam direction  \cite{momeni_hasan_abadi_ultra-wideband_2015}. To the best of our knowledge, there's no practical TTD-based RIS prototype until now. 

An alternative way to address the near-beam split is to form wideband beam with acceptive gain in the whole frequency band. Researchers in \cite{cheng_achievable_2024} partitioned the RIS into $N_\text{sub}$ virtual subarrays (VSA), configuring each subarray to focus its beam at the user's location at different frequencies. This approach aims to achieve relatively stable gain across all subcarriers, but the gain of each subarray is limited to $1/ N_\text{sub}^2$ compared to the entire RIS due to the proportional relationship between beamforming gain and the square of the number of RIS elements \cite{wu_towards_2020}.  Therefore, the separate beamforming design on subarrays can not get the benefit of the whole RIS plane, resulting in severe gain loss. Hence, an effective and practical beamforming method is required to deal with the near-field beam split effect of RIS.

\subsection{Contributions}
To fill in this gap, the characteristics of the Fresnel zone introduced by the RIS cascaded channel are revealed, and a near-field wideband RIS beamforming method based on the Fresnel zone without extra hardware cost is proposed to overcome the near-field beam split effect of RIS. The main contributions of this paper include:

\begin{itemize}
	\item{\cb	We investigate the Fresnel zone of RIS cascaded channel and propose a transition of coordinate system based on Fresnel zone. The Fresnel zones are a set of ellipsoidal surface with same two focus points at the transmitter and receiver. Phase of channels via RIS elements on the same Fresnel zone is aligned as their route lengths are identical. By designing the phase shifts of elements in the same Fresnel zone to be equal, the reflected signals from these elements combine in-phase at the receiver regardless of the frequency. This finding indicates that beam split does not occur when considering only RIS elements within one Fresnel zone. However, beam split persists across different Fresnel zones. Therefore, we reorient the coordinate system to two dimensions: along and across the Fresnel zones.  In the direction along the Fresnel zone, we design uniform phase shifts to focus signals on the user's location. Then, the effective channel is simplified from a \textit{two-dimensional} summation over the RIS plane with \textit{nonlinear} phase term to a \textit{one-dimensional} summation with \textit{linear} phase across different Fresnel zones. Further, we reveal that the equivalent channel over frequencies is simply a Fourier transform of the reflective intensity of each Fresnel zone, modulated by the designed phase, facilitating subsequent analysis and design. }

	\item{The equivalent channel gain over the frequency band influenced by the near-field beam split effect of RIS is analyzed based on the proposed Fourier transform relationship. Although prior works have identified severe beam gain loss on edge subcarriers due to beam split effect, the closed-form expression of channel gain has yet been given. With a basis in the observation that the channel across different frequency is the Fourier transform of channel intensity across Fresnel zones for classical narrowband beamforming, we propose a nearly closed-form expression to approximate the gain. Then, we analyze the 3dB bandwidth of equivalent channel with the concern of the beam split effect. With these results, we show that the effective bandwidth drops significantly as the aperture of RIS increases, leading to high performance loss. 
	}
	\item{The upper bound of achievable rate of phase shift-based wideband RIS-enabled communication system is proposed. As the gain is the spectrum of the phase-modulated intensity of Fresnel zones, the summation of gain across the overall frequency band $f\in \mathcal{R}$ is constant according to the Parseval's theorem. Therefore, the in-band power of channel gain is limited. With the power constraint, the achievable rate has an upper bound which is approached by ideal gain evenly distributed in frequency band with no out-of-band leakage at the receiver's location. Compared with upper bound proposed in  \cite{wang_wideband_2024}, which can only be reach with TTD architecture, the bound proposed in our work is tighter and more feasible for the phase-shift-based RIS.
	}
	\item{The optimization of achievable rate is reformulated as the minimization of the difference between the gain of designed beamforming and ideal gain. As the gain is the spectrum of the phase-modulated intensity of Fresnel zones, the phase design across Fresnel zones can be transformed into the design of a phase-modulated wave. Leveraging insights from radar system designs, we apply the stationary phase method on the design of phase-modulated wave to shape the desired spectrum effectively. {\cb Additionally, we introduce a Gerchberg-Saxton (GS) algorithm to further enhance the beamforming performance.} Finally, the phase design for RIS beamforming is completed with the design on both dimension of along and across the Fresnel zones. Simulation results show that the achievable rate of the proposed method approaches the upper bound, confirming the effectiveness of the proposed Fresnel zone based method in mitigating the near-field beam split effect. 	
}
\end{itemize}

\subsection{Organization and Notation}
The rest of this paper is organized as follows. We introduce the system model of the RIS-aided wideband communication and the near field region is analyzed in Section \ref{System_model}. The Fresnel zone is introduced in Section \ref{Fresnel}, with the coordinate transformation and the dimension reduction based on Fresnel zone. The near-field beam split effect is analyzed in Section \ref{split}. The upper bound of achievable rate for RIS-enabled communication system is given in Section \ref{upperbound}. The Fresnel zone based near-field wideband beamforming design is proposed in Section \ref{beamforming}. Simulation results are provided in Section \ref{result}, followed by the conclusions in Section \ref{conclusion} .

\textit{Notation}: Lower-case boldface letters $\mathbf{x}$ denote vectors, upper-case boldface letters $\mathbf{X}$ denote matrix and calligraphic letters $\mathcal{S}$ denote set; $(\cdot)^*, (\cdot)^T, (\cdot)^H$ and $\|\cdot\|^k$ denote the conjugate, transpose, conjugate transpose and k-norm of a vector or matrix  respectively; $\text{diag}(\mathbf{x})$ denotes a diagonal matrix whose $i\text{th}$ element element on diagonal is equal to the $i\text{th}$ entry of vector $\mathbf{x}$. $\mathcal{CN}(\mu, {\bm\Sigma}) $ and $\mathcal{U}(a,b)$ denote the Gaussian distribution with mean $\mu$ and covariance ${\bm\Sigma}$, and the uniform distribution between $a$ and $b$, respectively.

\section{System Model}
\begin{figure}[!t]
	\centering
	\includegraphics[width=0.45\textwidth]{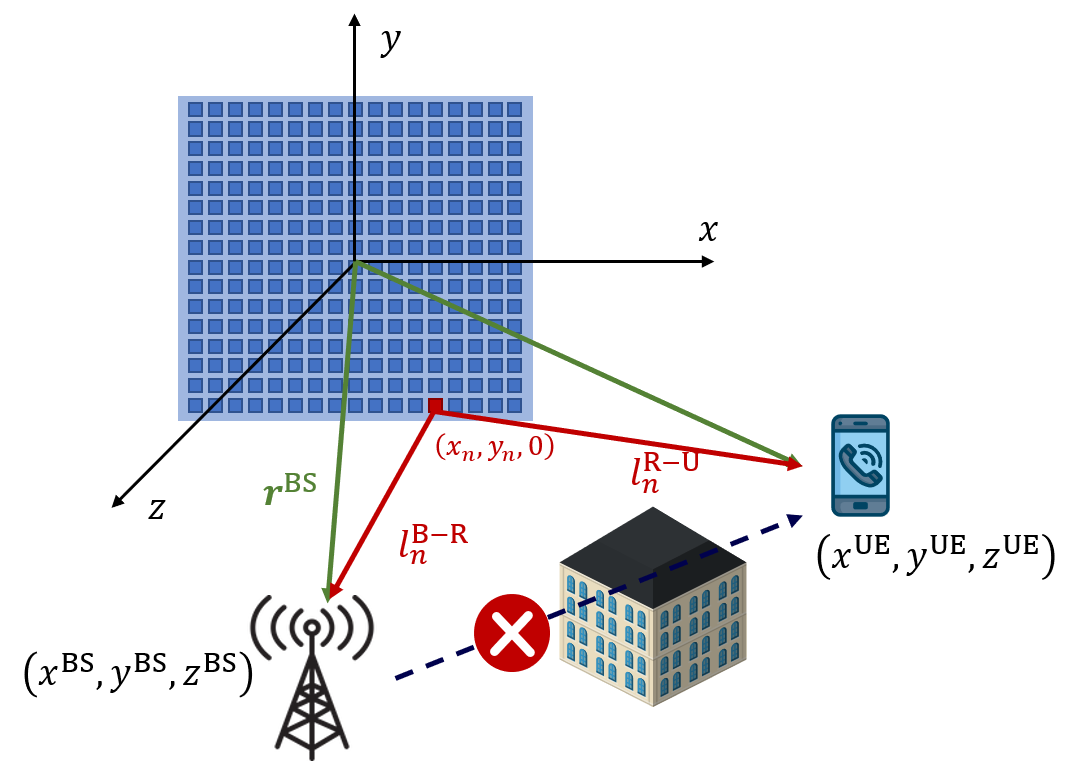}
	\caption{System model.}
	\label{fig:system_model}
\end{figure}

{\label{System_model}}
We consider an RIS-enabled wireless communication system over a bandwidth of $B$ around a carrier frequency of $f_c$, as shown in Fig. \ref{fig:system_model}. We consider a square RIS array with a side length of $D$. The number of subcarriers is $K$ and the frequency of  $k-th$ subcarrier is expressed as $f_k=f_c+B\left(\frac{2k-1}{2K}-\frac{1}{2}\right)$, for $k=1,...,K$. The RIS lies in the $xy$ plane centering at the origin $(0,0,0)$ and the edges of RIS are parallel to the axes, respectively. The RIS has $N^\text{RIS}_1$ and $N^\text{RIS}_2$ elements on $x$ and $y$ axis respectively, with a total number of $ N^{\mathrm{RIS}}=N^\text{RIS}_1N^\text{RIS}_2$.
 The space between successive elements along the $x$ and $y$ dimensions is $d$. 
 The set of 3D coordinates of elements of RIS is defined as $\mathcal{S}$, which has $N^\ris$ coordinates that satisfy $-D/2<x<D/2$, $-D/2<y<D/2$ and $z = 0$. 
 The coordinate of RIS element $(n_x,n_y)$ is $\mathbf{r}_n^\text{RIS}= \left((n_x-\frac{N_1^\ris+1}{2})d,(n_y-\frac{N_2^\ris+1}{2})d,0 \right) $, where the index of element $n$ is defined as $n = (n_x-1)N^\ris_2+n_y$. 
 The base station (BS) located at $\mathbf{r^\bs}=(x^\bs, y^\bs, z^\bs)$ is assumed to have $N^\bs=N^\bs_1N^\bs_2 $ antennas with $N^\bs_1$ and $N^\bs_2$ in the direction  $\mathbf{u}_1$ and $\mathbf{u}_2$ respectively. The location of BS antenna $(n_1,n_2) $ is $\mathbf{r}^\bs_n = \mathbf{r}^\bs+n_1 \mathbf{u}_1+n_2 \mathbf{u}_2$, where the index of element $n$ is defined as $n = (n_1-1)N^\bs_2+n_2$. 
 A single receiver (UE) is considered with the location at $\mathbf{r^\ue}= (x^\ue,y^\ue,z^\ue)$. 
 Then the distance from $n$-th RIS element at $\mathbf{r}_n$ to UE and $m$-th element of BS can be expressed as  $l^\ru_n=\|\mathbf{r}_n-\mathbf{r}^\ue\|_2$  and  $l^\br_{n,m}=\|\mathbf{r}_n-\mathbf{r}^\bs_m\|_2$respectively. Furthermore, $R^\br =\|\mathbf{r}^\bs\|_2$ and $R^\ru = \|\mathbf{r}^\ue\|_2$ are defined as the central distance.  

In this paper, the BS-UE LOS channel is assumed to be blocked. As the gains of NLOS channels are usually negligible in mmWave/THz band, only the BS-RIS and RIS-UE LOS channels are considered. The received signal of the $k$-th subcarrier can be expressed as
\begin{equation}
	y_k = \left({\mathbf{h}^\ru_k}\right)^T\bm{\Theta}\mathbf{H}^\br_k\mathbf{v}_kx_k+n_k
	\label{signal model},
\end{equation}
where, $x_k\in \mathbb{C}$ is the transmitted signal from the BS with power of $s$, {\cb$v_k\in \mathbb{C}^{N^\bs\times 1}$ is the precoding vector of BS, }and $ n_k\sim \mathcal{CN}(0,\sigma_n^2) $ is the additive write Gaussian noise. 
The RIS elements are assumed to follow the constant reflection amplitude constraint \cite{yang_1-bit_2016, dai_reconfigurable_2020} and the reflection coefficient matrix can be written as $\bm{\Theta}  =\mathrm{diag}(\mathbf{w})$, where {\cb$ \mathbf{w} = \left[\mathrm{e}^{\mathrm{j}\phi_1},\dots, \mathrm{e}^{\mathrm{j}\phi_{N^\ris}}\right] $} . 
As the near-field LOS channel model  are considered \cite{cui_near-field_2023}, the entries of RIS-UE channel vector $\mathbf{h}^\ru_k\in\mathbb{C}^{N^\ris\times 1}$ and BS-RIS channel matrix $\mathbf{H}^\br_k\in\mathbb{C}^{N^\ris\times N^\bs}$ can be written as 
\begin{subequations}
	\begin{align}
		\left[\mathbf{h}^\ru_k\right]_n &= \frac{c}{2\pi f_kl^\ru_n}e^{-j2\pi f_kl^\ru_n/c}\label{eq_hru}\\
		\left[\mathbf{H}^\br_k\right]_{(n,m)} &=  \frac{c}{2\pi f_kl^\br_{n,m}\sqrt[]{N^\bs}}e^{-j2\pi f_kl^\br_{n,m}/c}.\label{eq_Hbr}
	\end{align}
\end{subequations}

{\cb In this study, we assume that the array size of the BS is relatively small compared to that of the RIS. This is primarily due to the high hardware costs associated with the BS antenna system. In this context, we adopt a far-field assumption for the BS, indicating that the RIS is positioned in the far field of the BS. Conversely, the BS remains in the near field of the RIS.}  Under this assumption, distance term $l^\br_{n,m}$ in \eqref{eq_Hbr} is approximated by $l^\br_{n,m} = \|\bfr^\br_n-(\bfr^\br_{n,m}-\bfr^\bs)\|_2\approx l^\br_n-\xi_1m_1d-\xi_2m_2d$, where $l^\br_n$ is the distance from the center of BS to $n$-th RIS element, $\xi_1$ and $\xi_2$ are the angles-of-departure (AoD) at BS side and $m$ is decomposed to $m = (m_1-1) N^\bs_2+m_2$. Then the BS-RIS channel matrix can be rewritten as
\begin{equation}
	\mathbf{H}^\br_k = \bfh^\br_k\left({\bfh^\bs_k}\right)^T,
\end{equation}
where $\bfh^\br_k$ is the near-field array response at RIS with $	\left[\mathbf{h}^\br_k\right]_n = \frac{c}{2\pi fl^\br_n}e^{-j2\pi f_kl^\br_n/c}$ and $\bfh^\bs_k$ is the far-field array response at BS with $\left[\bfh^\bs_k\right]_m = \frac{1}{\sqrt{N^\bs}} e^{j2\pi f_kd(\xi_1m_1+\xi_2 m_2)/c}$. Then the system model can be written as
\begin{equation}
	y_k = \mathbf{w}^T\bfh^\C_k\left({\bfh^\bs_k}\right)^T\mathbf{v}_k x_k+n_k, 
\end{equation}
where $\bfh^\C_k = \bfh^\ru_k\odot\bfh^\br_k$ denotes the cascaded channel of RIS. 

As our work is focused on the beamforming of RIS, the beamforming of BS is assumed to be ideal, i.e. $\mathbf{v}_k = \left({\bfh^\bs_k}\right)^*$. Then the equivalent channel from the BS to the UE can be expressed as
\begin{equation}
	{g}_k =\mathbf{w}^T\bfh^\C_k\left({\bfh^\bs_k}\right)^T\mathbf{v}_k= \sqrt{N^\bs} {\mathbf{w}}^T\mathbf{h}^\C_k.
	\label{g}
\end{equation}


\begin{figure}[!t]
	\centering
	\includegraphics[width=0.45\textwidth]{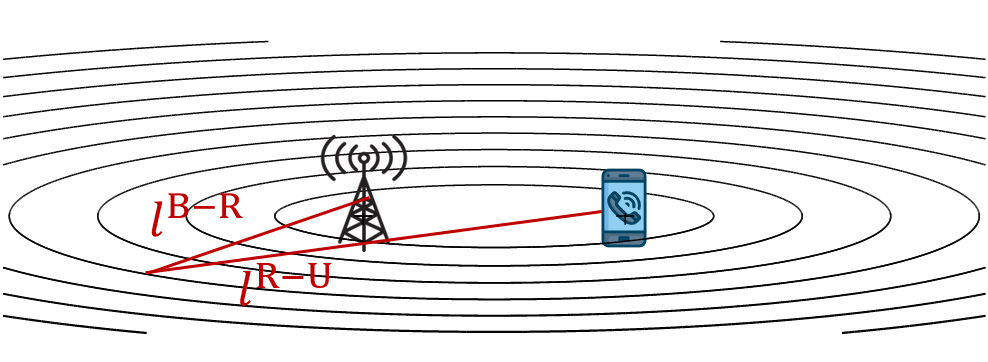}
	\caption{A series of Fresnel zones of the communication system with a BS and a UE. The transmission length $l^\br+l^\ru$ remains the same on each Fresnel zone.
	}
	\label{fig:Fresnel_zones}
\end{figure}

\section{Fresnel Zone Model for RIS Channel}{\label{Fresnel}}
The equivalent channel $g_k$ in \eqref{g} is the summation of element-wise cascaded channels via elements over the two-dimensional RIS plane, with non-linear distance phase terms, making the analysis and beamforming design complex. In this section, the concept of Fresnel zone is introduced to deal with the RIS cascaded channel. Then we will show that RIS elements on a single Fresnel zone will not suffer from beam split effect, followed by a coordinate transformation which can reduce the problem to a one-dimensional integration across Fresnel zones with linear phase term. 

\subsection{Introduction to the Fresnel Zone}
In the context of radio propagation, Fresnel zones refer to the concentric ellipsoids whose foci are the transmitter and receiver. The property of ellipsoids ensures that the signals reflected by RIS elements on the same Fresnel zone have the same route length $l = l^\br+l^\ru$, as shown in Fig. \ref{fig:Fresnel_zones}. 

The intersection of ellipsoids of Fresnel zones and the RIS plane are a series of ellipses. It's hard to determine correspondence between the RIS elements and the Fresnel zones, due to mismatch of continuous-shaped Fresnel zones and the discrete spacing of RIS. To address this, we use an imaginary continuous RIS (I-RIS) to approximate the RIS with discrete spacing. The I-RIS contains an uncountable number of virtual elements lying in the set $\mathcal{S}_C = \{(x,y)||x|<D/2, |y|<D/2\}$.  The overall reflection gain of the virtual elements within the area of an original RIS element should equal the gain of that RIS element. Therefore, the weight of the virtual element located at $(x,y,0)$ is normalized to $w(x,y)=e^{j\phi(x,y)}/d^2$. As the elements of the discrete RIS can be regarded as spatial sampling of continuous I-RIS, the accuracy of the approximation can be ensured by the Nyquist spatial sampling theorem. Similar to the expression in \eqref{g}, the equivalent channel for I-RIS system is expressed as   
\begin{equation}
	g(f) = g_0\int_{-\frac{D}{2}}^{\frac{D}{2}}{\int_{-\frac{D}{2}}^{\frac{D}{2}}{e^{j\phi-j2\pi f(l^\br+l^\ru )/c} \mathrm{d} x }\mathrm{d} y},
	\label{g_int}
\end{equation}
where $g_0= \frac{\sqrt{N^\bs}c^2}{4\pi^2 f^2R^\br R^\ru d^2}$ is the basic path-loss. Here we write the equivalent channel $g(f)$ as a function of frequency to simplify presentation. {\cb Besides, we use $R^\br$ and $R^\ru$ to approximate $l_n^\br$ and $l_n^\ru$, respectively. Since the aperture is significantly less than the transmit distance, the approximation in amplitude has mere impact on the beamforming performance \cite{cui2022channel}.}

The idea of continuous spacing I-RIS is just a mathematical technique to aid the analysis and beamforming design as it fits the continuity of Fresnel zones. Although they share same concept of continuous surface, the I-RIS should not be confused with holographic metasurface, because their models are quite different. { \cb Holographic metasurfaces predominantly focus on electromagnetic field theory, incorporating Maxwell's equations and boundary conditions \cite{gong2024holographic,Dardari2020Communicating}, in contrast to the simpler approximation utilized in this context. } Our further analysis on I-RIS can be regarded as a close approximation of the discrete RIS, while the phase design for the I-RIS will be sampled at the points where elements of the discrete RIS are located.  
 
\subsection{Beam Split Effect on Fresnel Zones}
The cause of beam split is the mismatch of \textit{frequency-independent} phase shifts and \textit{frequency-dependent} channel, as the whole RIS is considered. To be specific, the phase shifts of RIS are designed to compensate the phase of the channel at the center frequency, i.e. $\phi = 2\pi fl/c$, to add the signals in phase at the receiver, achieving maximum gain. As the frequency increases, the \textit{frequency-independent} phase shifts of RIS keep the same, while phase of element-wise cascaded channel $-2\pi fl/c$ \textit{varies differently} across different elements. The phase of signal is not aligned, resulting in a loss of gain. 

Different from the whole RIS, beam split effect disappears when only virtual I-RIS elements on one Fresnel zone are considered. Specifically, channels via elements on the Fresnel zone have \textit{aligned} phase $-2\pi fl/c$ as they share the same route length $l$, realizing the in-phase mixture of signal at the receiver. As the phase of these elements are designed to be the same, the gain remains maximized and the beam remains focused to the UE's location regardless of frequency. Therefore, elements on one Fresnel zone do not suffer from beam split effect.

{\cb When we return to the discussion of the entire UPA RIS, beam splitting still occurs due to the presence of multiple Fresnel zones. In the upcoming sections, we will categorize the I-RIS elements into their respective Fresnel zones and conduct analysis of the beam splitting effects across these different zones.}

\subsection{Coordinate Transformation Based on Fresnel Zone}
In this subsection, we transform the Cartesian coordinate to the proposed Fresnel-zone coordinate with two dimensions of along and across the Fresnel zones on the I-RIS plane.

The Fresnel zones of the system are a set of concentric ellipsoids whose focus points are the transmitter and receiver. The semi-major axis length of the ellipsoid is defined as $a$. According to the property of ellipsoids, the length of reflecting route is $l = l^\br+l^\ru=2a$. Therefore, we can get the ellipsoid function of Fresnel zone
\begin{equation}
	\begin{aligned}
		2a = &\sqrt{(x^\bs-x)^2+(y^\bs-y)^2+(z-z^\bs)^2}\\&+\sqrt{(x^\ue-x)^2+(y^\ue-y)^2+(z-z^\ue)^2} 
	\end{aligned}
	\label{ellipsoid}
\end{equation}

The intersection of the Fresnel zone shown in \eqref{ellipsoid} and the plane $z = 0$ where the RIS is located is a ellipse. The function of the intersection ellipse can be written as
\begin{equation}
	\begin{aligned}
		&\sqrt{(x^\bs-x)^2+(y^\bs-y)^2+{z^\bs}^2}\\&+\sqrt{(x^\ue-x)^2+(y^\ue-y)^2+{z^\ue}^2} = 2a.
	\end{aligned}
	\label{ellipse0}
\end{equation}

\begin{figure}[t]
	\centering
	\includegraphics[width=0.9\linewidth]{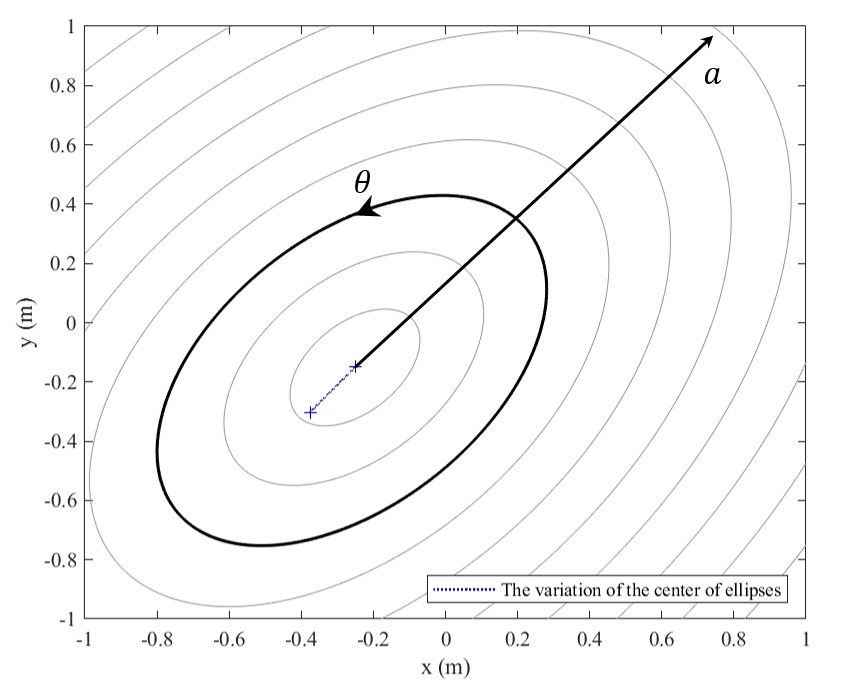}
	\caption{The intersections of Fresnel zones and the RIS plane are a series of ellipses. A new coordinate system is set on the ellipses with axes of semi-major axis $a$ of Fresnel zone and angular coordinate $\theta$  }
	\label{fig:ellipse}
\end{figure}

In order to simplify the subsequent process, coordinate translation and rotation are adopted so that the two roots in \eqref{ellipse0} have symmetry format. Specifically, we establish a new Cartesian coordinate system. The origin is the projection of the midpoint of the connection between the transmitter and the receiver on the RIS plane, the x-axis direction is the direction of the transmitter pointing to the receiver, the z-axis direction remains unchanged, and the y-axis direction is obtained from the right-hand rule of the Cartesian coordinate system. The coordinate translation can be written as
\begin{equation}
	\begin{cases}
		x' = (x-x_c)\cos{\alpha}+(y-y_c)\sin{\alpha}\\
		y' = -(x-x_c)\sin{\alpha}+(y-y_c)\cos{\alpha},
	\end{cases}
	\label{rotation}
\end{equation}
where $x_c = \frac{x^\bs+x^\ue}{2}, y_c =\frac {y^\bs+y^\ue}{2}$are the middle point of transmitter and receiver, $\alpha=\arctan{\frac{y^\bs-y^\ue}{x^\bs-x^\ue}}$ is the angle of rotation. Then the ellipse in \eqref{ellipse0} can be expressed as
\begin{equation}
	\sqrt{(x'+u)^2+y'^2 +{z^\bs}^2 }+\sqrt{(x'-u)^2+y'^2 +{z^\ue}^2 }=2a
	\label{ellipse1},
\end{equation}
where $u=\frac{1}{2}\sqrt{(x^\bs-x^\ue)^2+(y^\bs-y^\ue)^2}$ is half of the projection length of transmission route to the plane of RIS, and $b = \sqrt{a^2-u^2}$. Then we express the ellipse in \eqref{ellipse1} into the standard form
\begin{equation}
	{(x'-x_0)^2\over a^2\eta_0^2}+\frac{y'^2}{b^2\eta_0^2}=1,
	\label{eq:ellipsestd}
\end{equation}
where $x_0 = {1\over 4}u{{z^\bs}^2-{z^\ue}^2\over(a^2-u^2)^2}$is the center of the ellipse, and $\eta_0 = \sqrt{1+{\left({z^\ue}^2-{z^\bs}^2\right)^2\over 16b^4}-{1\over 2}{{z^\bs}^2+{z^\ue}^2\over b^2}}$ is the correction factor for the semi-major axis and the semi-minor axis, and the length of the semi-major axis and the length of the semi-minor axis become $\eta_0$ times of the original. As $a$ changes, the center of the ellipse changes. The series of ellipses are drawn in Fig. \ref{fig:ellipse}.

In classical polar coordinate system, the Cartesian coordinates $x$ and $y$ can be converted to the radial coordinate $r$ and the angular coordinate $\theta$. In the Fresnel zone system,  we use the length of semi-major axis $a$ to replace the radial coordinate $r$, while the angular coordinate $\theta$ remain the same, as shown in Fig. \ref{fig:ellipse}. Then the transformation of coordinate can be expressed  by using the trigonometric functions  
\begin{equation}
	\begin{cases}
		x' & = a\eta_0\cos\theta+x_0\\
		y' &= \sqrt{a^2-u^2}\eta_0 \sin\theta.
	\end{cases}
	\label{eq:xy_atheta}
\end{equation}

After the transformation of coordinate, the point set of I-RIS elements changes from $\mathcal{S}_C$ to $\mathcal{V}_0=\left\{(a,\theta)|\left(x(a,\theta),y(a,\theta),0\right)\in \mathcal{S}_C\right\}$, where $x(a,\theta)$ and $y(a,\theta)$ are the original Cartesian coordinates of the point $(a,\theta)$ in the Fresnel zone coordinate. Then the equivalent channel can be rewritten as
\begin{equation}
	g(f) = \int_{a}{e^{-j2\pi f\frac{2a}{c}}\int_{\theta}g_0e^{j\phi(a,\theta)}J(a,\theta) \mathrm{d}\theta}\mathrm{d} a,
	\label{hbb1}
\end{equation}
where $J(a,\theta) = \left| \frac{\partial(x',y')}{\partial (a,\theta)}  \right|$ is the absolute value of the Jacobian of the transformation. It can be further expressed as 
\begin{equation}
	\begin{aligned}
		J(a,\theta) &= \left |\mathrm{det}{\partial(x,y)\over \partial (a,\theta)}\right | \\&= -{au({z^\ue}^2-{z^\bs}^2)\over b^3}\eta_0 \cos\theta 
		\\&	+ {a^2\over2}\left(  -{({z^\ue}^2-{z^\bs}^2)^2\over 4b^5}+{{z^\bs}^2+{z^\ue}^2\over b^3}   \right)\\&~~+{\eta_0^2(b^2+{1\over2}u^2)\over b }-{\eta_0^2u^2\over 2b}\cos{2\theta}.
		\label{eq:Jacobi}
	\end{aligned}
\end{equation}

Note that the inner integral along the Fresnel zone in \eqref{hbb1} is frequency independent, so the beam split effect will not occur when the I-RIS elements on one Fresnel zone is considered. We can design the phase design $\phi(a,\theta)$ to be the same on the Fresnel zone with semi-major length $a$, i.e. $\phi(a,\theta) = \psi(a)$, so that the reflective intensity of each Fresnel zone can always reach its maximum value.

The inner integral then becomes a real value integral of the Jacobian 
\begin{equation}
	v(a) = \int_{\theta\in \mathcal{V}(a)}g_0 J(a,\theta)\text{d}\theta,
	\label{va}
\end{equation} 
where $\mathcal{V}(a) = \{\theta|(a,\theta)\in \mathcal{V}_0\}$ is the set of  angle of point which lays in the region of RIS. The value of integral $v(a)$ is regarded as the reflective intensity of Fresnel zone with semi-major length $a$. Observing that the Fresnel zone may be cut by the edge of RIS to several pieces, so $\mathcal{V}(a)$ is a union of several continuous intervals. The endpoints of the intervals can be obtained by solving the quadratic equation combining the ellipse function of Fresnel zone and the line function of RIS edge. After determining the intervals of the integral, its value will be easily solved since $J(a,\theta)$ is simply a polynomial of $\cos(\theta)$ and $\sin(\theta)$. After we get the value of the inner integral $v(a)$,  the equivalent channel can be rewritten as 
\begin{equation}
	g(f) = \int_{a}
	v(a)e^{j\psi(a)}
	e^{-j2\pi f\frac{2a}{c}} \mathrm{d} a.
	\label{hbb1d}
\end{equation}

\begin{figure*}[t]
	\centering
	\subfloat[]{\includegraphics[width=0.33\linewidth]{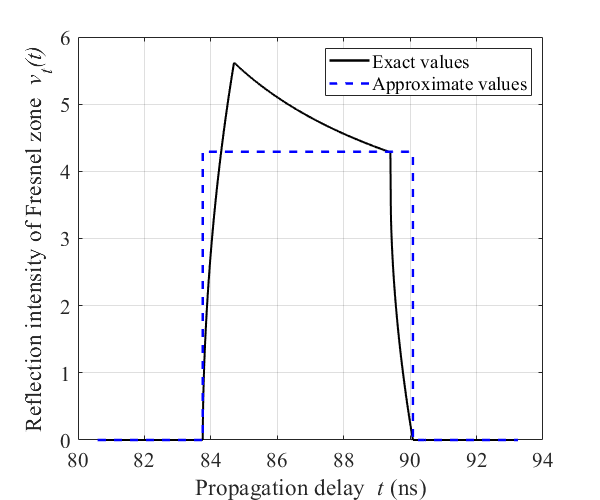}%
		\label{fig:intensity}}
	\hfil
	\subfloat[]{\includegraphics[width=0.33\linewidth]{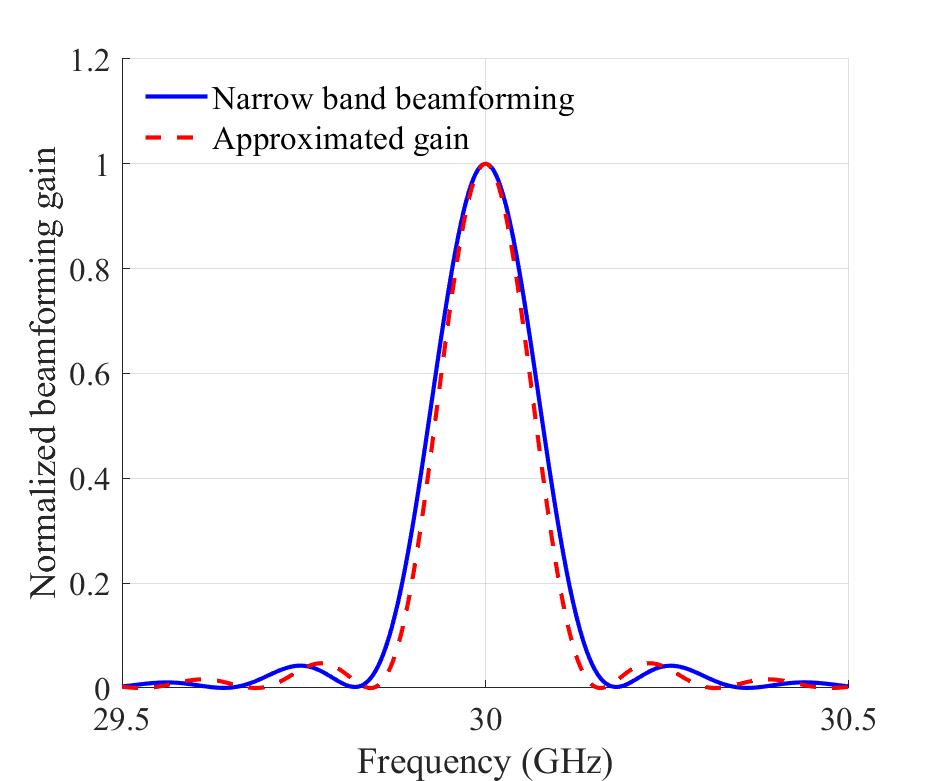}%
		\label{fig:sinc}}
	\subfloat[]{\includegraphics[width=0.33\linewidth]{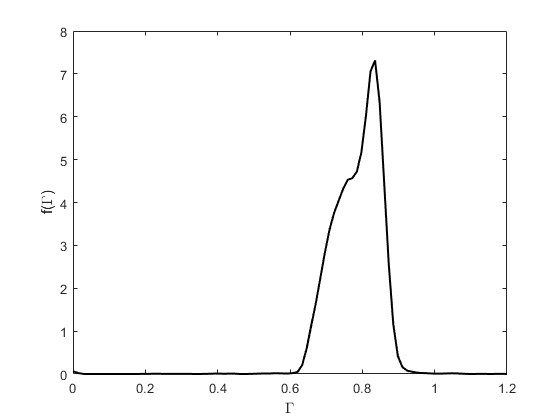}
		\label{fig_pdfgamma}}
	\caption{The approximation of channel gain $|g_\text{Narr}(f)|^2$. The approximation of $v_t(t)$ is shown in Fig. \ref{fig:intensity}. The approximation of channel gain is illustrated in Fig. \ref{fig:sinc}.  	Probability density function of $\Gamma $ as the location of BS and UE chosen randomly is shown in Fig. \ref{fig_pdfgamma}. }
	\label{fig:beam_profiles0}
\end{figure*}

Here we get a one-dimensional integral across the Fresnel zones, with phase term $-j2\pi f\frac{2a}{c}$ linear to the integral variable $a$. Further, the integral across the Fresnel zones can be regarded as a spatial Fourier transform. If we use the propagation delay $t = 2a/c$ to replace $a$, the integral becomes 
\begin{equation}
	g(f) = \int_{t}
	v_t(t)e^{j\psi_t(t)}
	e^{-j2\pi ft} \mathrm{d} t,
	\label{ftransform}
\end{equation}
 where we use {\cb$v_t(t)=\frac{c}{2} v\left(t\frac{c}{2}\right)$} and $\psi_t(t) = \psi(tc/2)$ to simplify writing. Therefore, the channel is the Fourier transform of Fresnel zone intensity $v_t(t)$ modulated by phase design $\psi_t(t)$. 

\section{Near-field Beam Split of RIS}{\label{split}}
In this section, the near-field beam split effect is analyzed and the gain loss with respect to the bandwidth and the deployment of RIS is analyzed.

\begin{figure}[!t]
	\centering
	\includegraphics[width=0.5\textwidth]{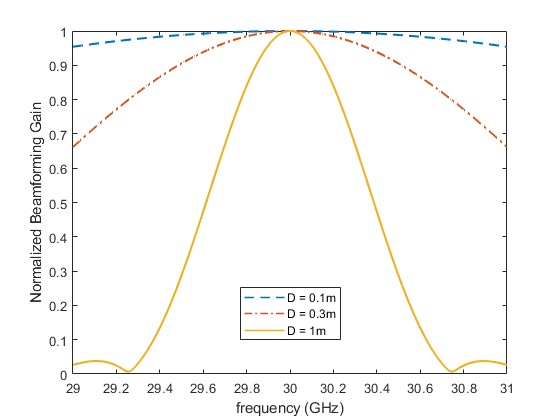}
	\caption{ Normalized beamforming gain with respect to frequency. With increase of aperture $D$, the effective bandwidth drops significantly.}
	\label{fig_NFBM}
\end{figure}

It's quite easy to analyze the near-field beam split effect through the Fourier transform relationship of equivalent gain $g(f)$  and the intensity of Fresnel zone modulated by the designed phase, as expressed in \eqref{ftransform}. Specifically, the system suffers from the near-field beam split as the near-field narrowband beamforming $\psi_t(t) =  \exp(j2\pi f_ct/c)$ is adopted, with equivalent channel of 
\begin{equation}
	g_\text{Narr}(f) = V(f-f_c).
	\label{ftransform1}
\end{equation}
where $V(f) = \int_{t}v_t(t)e^{-j2\pi ft} \mathrm{d} t$ is the Fourier transform of $v_t(t)$.  As $v_t(t)$ is a positive-valued function, the channel gain $|g_\text{Narr}(f)|^2$ can reach its maximum at the center frequency $f_c$. When the deviation of frequency increases, the gain will drop significantly, as illustrated by the numerical results in Fig. \ref{fig_NFBM}. This phenomenon is called the beam split effect. The system  suffers from more decrease in gain as the size of RIS increases.  

Further, we give an approximation to $g_\text{Narr}(f)$ in order to establish the mathematical relationship between RIS aperture and the 3dB bandwidth of gain. Observing that $v_t(t)$ is positive with little fluctuation, as shown in Fig. \ref{fig:intensity}, we can adopt its zero-order approximation 
\begin{equation}
	\hat{v}_t(t) = \frac{\int v_t(t) \text{d}t}{t_\text{max}-t_\text{min}} \stackrel{(a)}{=}  \frac{g_0N^\ris}{t_\text{max}-t_\text{min}},
	\label{hatv}
\end{equation}
where $t_\text{max}$ and $t_\text{min}$ are the maximum and the minimum propagation delay of paths via elements on the I-RIS, respectively. Equation $(a)$ in \eqref{hatv} holds because $\int{v}_t(t)\text{d}t = V(0) = g_\text{Narr}(f_c) = g_0N^\ris$.  The approximated channel can then be written as
\begin{equation}
	\hat{g}_\text{Narr}(f) = g_0N^\ris sinc((f-f_c)\Delta_t), 
\end{equation}
where $sinc(x) = \frac{\sin (\pi x)}{\pi x}$ is the sinc function and $\Delta_t = t_\text{max}-t_\text{min}$. As shown in Fig. \ref{fig:sinc}, the gain of approximated channel closely aligns with the exact value in the main lobe,  highlighting the effective approximation performance.  According to the property of sinc function, the 3dB bandwidth of $g(f)$ is approximated by
\begin{equation}
{B}_\text{3dB} \approx	\hat{B}_\text{3dB} = \frac{\Gamma_0}{\Delta_t},
\label{NFBM_2}
\end{equation}
 where $\Gamma_0 = 0.886$ is the constant to determine the 3dB bandwidth of sinc function \cite{oppenheim_discrete-time_1999}. Now we expand $t_\text{max}$ around the point $\mathbf{r}$ where $t$ get the minimum value $t_\text{min}$. Then $t_\text{max}$ can be approximated by
\begin{equation}
	\begin{aligned}
		t_\text{max}  \approx t_\text{min} +(\mathbf{r}_{a}-\mathbf{r}_{i})\nabla t \left. \right|_{\mathbf{r}=\mathbf{r}_i}
		\le t_\text{min}+ \iota D_0/c,
	\end{aligned}
	\label{NFBM_3}
\end{equation}
where $\iota = \sqrt{\left(\frac{x^\bs}{R^\br}+\frac{x^\ue}{R^\ru}\right)^2+\left(\frac{y^\bs}{R^\br}+\frac{y^\ue}{R^\ru}\right)^2     } $ is the scale factor corresponding to the direction of the BS and UE. It can be shown that $\iota$ is larger when the BS and the UE are in the similar direction and is smaller when they are in the opposite direction. 
Based on \eqref{NFBM_2} and \eqref{NFBM_3}, $\hat{B}_\text{3dB}$ can be expressed as
\begin{equation}
	 \hat{B}_\text{3dB}=\frac{c\Gamma_0}{\iota D_0}.
	\label{NFBM_4}
\end{equation}

To verify the result in \eqref{NFBM_4}, we simulate on the exact value of the factor $\Gamma = B_\text{3dB}\iota D_0/c$. The value of $\Gamma$ varies as the location of BS and UE changes. Our simulations showed that $\Gamma$ is approximately equal to the theoretical value of $\Gamma_0 = 0.866$, as illustrated in Fig. \ref{fig_pdfgamma}. This confirms the accuracy of our approximation of $\hat{B}_\text{3dB}$ in \eqref{NFBM_4}.

Therefore, the near-field beam split effect will be more significant when the aperture of RIS and bandwidth get larger and the direction of BS and UE get similar. We further provide the numerical result of the normalized beamforming gain in Fig. \ref{fig_NFBM}. The loss in the normalized beamforming gain increases with the increase of the aperture size $D_0$. With the RIS side length of 1m, the classical narrowband beamforming suffers more than 70\% gain loss in half of the band. At the center frequency, the beamforming gain of RIS increases proportional to the square of number of RIS elements. However, it will not increases as fast in the whole bandwidth due to the near-field beam split effect, which will limit the growth of system performance. 

\begin{figure}[!t]
	\centering
	\subfloat[]{\includegraphics[width=0.45\textwidth]{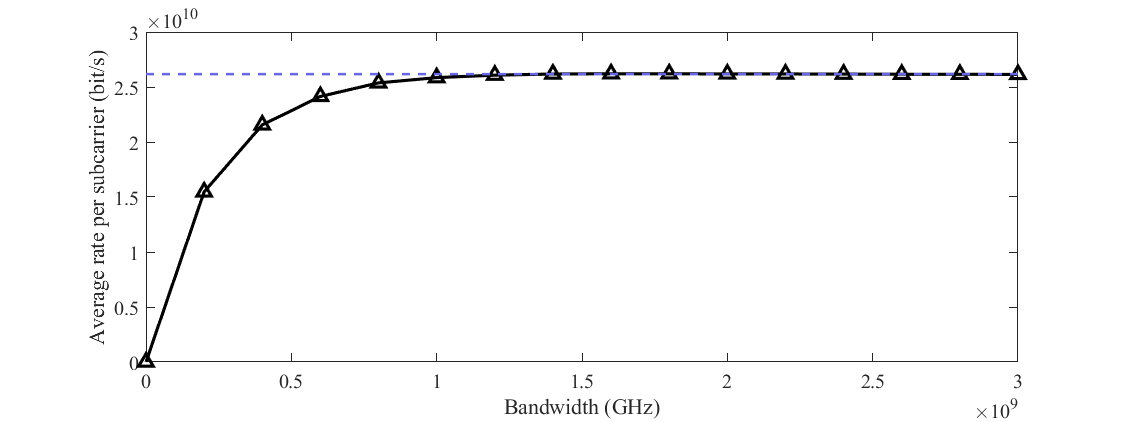}\label{fig:performance_bandwidth}}
	\hfil
	\subfloat[]{\includegraphics[width=0.45\textwidth]{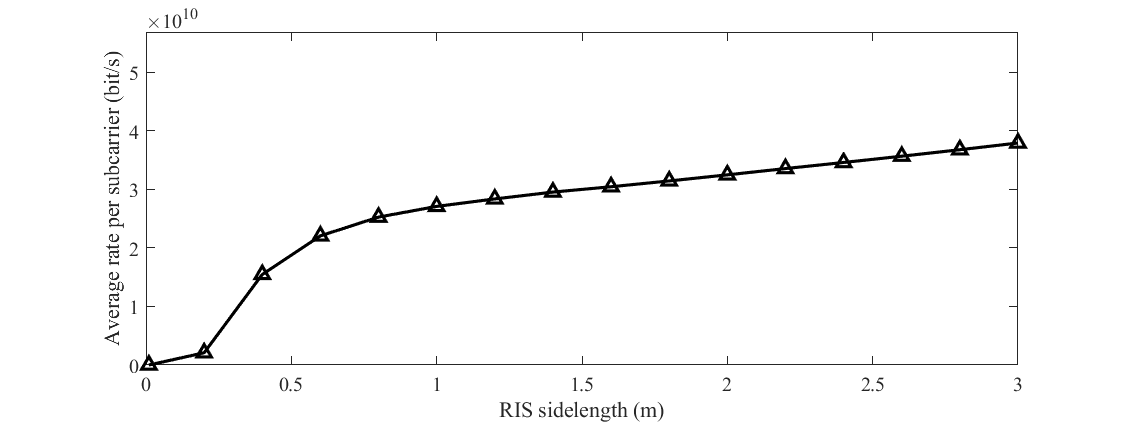}\label{fig:performance_D}}
	\caption{The limit in growth of achievable rate performance caused by near-field beam split effect.}
	\label{fig:beam_split_performance}
\end{figure}

The near-field beam split effect also imposes limitations on the potential increase of achievable rates with both the expansion of the RIS size and system bandwidth. This is evidenced by the fact that, as the bandwidth expands, the gain associated with classical beamforming approaches nearly zero within the supplementary bandwidth, rendering signal transmission in this range unfeasible. This phenomenon acts as a barrier, causing a bottleneck in achievable rates as bandwidth increases, as depicted in Fig. \ref{fig:performance_bandwidth} . Additionally, as the RIS size increases, the effective bandwidth of the gain decreases, restricting transmission to a handful of subcarriers around the center frequency. Although the gain at the center frequency experiences growth, its impact on achievable rates diminishes since rates are logarithmic functions of channel gains in high SNR scenarios. Consequently, the achievable rates show marginal improvements with the expansion of the RIS size, as illustrated in Fig. \ref{fig:performance_D}.

\section{Upper Bound of Achievable Rate in Wideband RIS System}{\label{upperbound}}
In this section, an upper bound of achievable rate is proposed for wideband RIS-enabled system. The result shows that the upper bound is reached when the channel gain distributes evenly in-band without out-of-band leakage.  

In this section, the number of subcarriers is assumed to be infinite with continuous frequency band. This assumption renders the result general,  as any subcarrier configuration can be seen as sampling of the continuous frequency band. The achievable rate can be then written as
\begin{equation}
	R = \int_{f_c-B/2}^{f_c+B/2}{\log_2\left(1+\frac{\left|g(f)\right|^2S_x}{S_\sigma}\right)\mathrm{d} f},
	\label{rate}
\end{equation}
where $S_x$ is the energy of signal and $S_\sigma$ is the power spectrum density of noise. 

The achievable rate is bounded by
\begin{equation}
	\begin{aligned}			&\int_{f_c-\frac{B}{2}}^{f_c+\frac{B}{2}}\log_2   \left(1+\frac{\left|g(f)\right|^2S_x}{S_\sigma}\right)\mathrm{d} f
		\\ \stackrel{(a)}{\le} &\log_2\left(1+\frac{S_x}{S_\sigma}  \int_{f_c-\frac{B}{2}}^{f_c+\frac{B}{2}}{\left|g(f)\right|^2 \mathrm{d} f} \right)
		\\ \stackrel{(b)}{\le} &\log_2\left(1+\frac{S_x}{S_\sigma}  \int_{-\infty}^{+\infty}{\left|g(f)\right|^2 \mathrm{d} f} \right)
		\\ \stackrel{(c)}{=}&\log_2\left(1+\frac{S_x}{S_\sigma}  g_t^2 \int_t{ v_t^2(t) \mathrm{d} t} \right),
	\end{aligned}
\label{Jensen}
\end{equation}
where (a) holds due to the Jensen's inequality and (c) is based on the Fourier transform relationship between $g(f)$ and $v_t(t)e^{j\psi_t(t)}$ in \eqref{ftransform} and the Parseval's theorem \cite{oppenheim_discrete-time_1999}. Here,  $E_g = \int_t{ v_t^2(t) 
\mathrm{d} t}$ is considered as the energy of channel, which is maximized as $v_t(t)$ has already been the highest reflective intensity along the Fresnel zone. Therefore, $\log_2\left(1+\frac{S_x}{S_\sigma}  E_g \right)$ is the upper bound of achievable rate for any phase configuration of the I-RIS. 

Additionally, we analyze the condition for the achievable rate to reach its upper bound. The equality in inequality (a) is attained when $|g(f)|^2$ remains constant in the frequency band $[f_c-B/2, f_c+B/2]$ and the equality in inequality (b) is achieved when no gain leaks out of the frequency band, i.e. $g(f) = 0, f\in (-\infty,f_c-B/2)\cup (f_c+B/2,+\infty)$. Therefore, the ideal gain to reach the upper bound can be written as
\begin{equation}
	\hat{g}(f)=\begin{cases}
		\sqrt{ \frac{E_g}{B}} & f \in [f_c-B/2, f_c+B/2]\\
		0 & \text{otherwise}.
	\end{cases} 
	\label{ideal}
\end{equation}

Note that the upper bound proposed here suits for the continuous I-RIS with infinite number of subcarriers. Real systems with discrete spaing RIS and finite number of subcarriers can be seen as the sampling in both space and frequency domain. The proposed upper bound and ideal condition still holds in real systems, as no benefit in rate can get in the sampling procedure. Besides, the ideal gain is not reachable, because the band-limited spectrum of gain can not be formed by a time-limited signal $v_t(t)e^{j\psi_t(t)}$. In the next section, we will propose phase design for $\psi_t(t)$ to approach the ideal gain spectrum.

\section{Wideband Beamforming Based on Fresnel Zone}{\label{beamforming}}
In this section, Fresnel-zone-based near-field wideband beamforming is proposed to mitigate the loss caused by near-field beam split. As it's hard to directly optimizing the achievable rate in \eqref{rate}, an alternating way is to design beams to approach the ideal spectrum that the gain distributes uniformly in the frequency band without out-of-band leakage. Specifically, the phase design of different Fresnel zone will be reformulated to phase design of frequency-modulation wave, which can be realized based on stationary phase method. Furthermore, we introduce the GS algorithm to enhance the overall performance. Finally, the phase design of one-dimensional phase modulation wave is mapped back to phase shift of RIS.

\subsection{\cb RIS Beamforming Based on Stationary Phase Method}
{\cb
Now we propose phase design for $\psi_t(t)$ in \eqref{ftransform} to make $|g(f)|$ approach  the ideal spectrum $\hat{g}(f)$ in \eqref{ideal}. Recall that in \eqref{ftransform}, $g(f)$ is the Fourier transform of a modulated wave $v_t(t)e^{j\psi_t(t)}$, where $v_t(t)$ is fixed and $\psi_t(t)$ is the modulating phase to be designed. Stationary phase (SP) method is a widely used approach for phase-modulated signal design in the field of radar, which can be adopted to approach the desired spectrum \cite{cook_radar_2012}. The stationary point $t_f$ of frequency $f$ satisfies with
\begin{equation}
	2\pi f = \psi_t'(t_f), 
	\label{eq:sp}
\end{equation}
where $\psi_t'(t)$ is the derivative of $\psi_t(t)$. Then the relation between the amplitude in frequency domain $g(f)$ and the amplitude in time domain $v_t(t_f)$ at the stationary point $t_f$ is approximately set up, with respect to the designed phase $\psi_t(t)$. Discussed with detail in \cite{cook_radar_2012}, this relation is given that
\begin{equation}
	P(t_f) = Q(f), 
	\label{eq:pq}
\end{equation} 
where  $P(t)$ and $Q(f)$ are defined as
\begin{equation}
	\begin{aligned}
		P(t) &= \int_{-\infty}^t v_t^2(\tau)\mathrm{d} \tau\\
		Q(f) &=  \int_{-\infty}^f |\hat{g}(f)|^2\mathrm{d} f.
	\end{aligned}
	\label{pq}
\end{equation}
Substitute $\hat{g}(f)$ in {\cb\eqref{ideal}} into the expression of $Q(f)$ we get
\begin{equation}
	Q(f) = \begin{cases}
		0& f<f_c- B/2\\
		\frac{E_g}{ B}(f-f_c+B/2)& f_c-B/2<f<f_c+B/2 \\
		E_g & f>f_c+B/2.
	\end{cases}.
\end{equation}
From \eqref{eq:pq} and \eqref{eq:sp}, we can get the derivative of the designed phase as
\begin{equation}
	\psi_t'(t_f) = 2\pi f = Q^{-1}\left(P(t_f)\right).
	\label{eq:psip}
\end{equation}
Now we reverse the relationship between $t_f$ and $f$ presented in \eqref{eq:sp}, presented $f$ is as a function of $t_f$ at the stationary point. For each given $t_f$, there must exist a corresponding $f$. Therefore, the relation in \eqref{eq:psip} applies for any $t_f\in [t_{\rm min}, t_{\rm max}]$. By integrating both the leftmost and rightmost sides of \eqref{eq:psip}, the phase profile can be appropriately designed as
\begin{equation}
	\begin{aligned}
		\psi_t(t) &= 2\pi \int_{-\infty}^t Q^{-1}\left[P(\tau)\right]\mathrm{d} \tau+ C \\
		&=2\pi \int_{t_{\rm min} }^t  \frac{BP(\tau)}{E}+f_c-\frac{B}{2}\mathrm{d} \tau+ C \\
		& = 2\pi B\int_{t_{\rm min} }^t \frac{P(\tau)}{E_g} d\tau +2\pi (f_c-\frac{B}{2})(t-t_{\rm min})+C
		,
	\end{aligned}
	\label{final_phase}
\end{equation}
where $C$ is some constant, which is set to $2\pi (f_c-\frac{B}{2})t_{\rm min}$ to simplify expression.

Therefore, we get the Fresnel zone domain phase shifts $\psi(a)=\psi_t(2a/c)$. Substituting $P(t)$ and $E_g$ into \eqref{final_phase}, we give a nearly closed-form expression of the design of phase shifts on Fresnel zones, which can be expressed as
\begin{equation}
	\psi(a) =  \frac{4\pi B}{c} \frac{\int_{a_\text{min}}^{a} \int_{a_\text{min}}^{a_1}v^2(a_2)\mathrm{d}a_2\mathrm{d}a_1}{\int_{a_\text{min}}^{a_\text{max}}v^2(a_2)\mathrm{d}a_2}+\frac{4\pi(f_c-B/2) a}{c}.
	\label{final_phase1}
\end{equation}

Finally, the phase of each RIS element can be calculated. The phase shift configuration of each RIS element gets the value of phase design of the Fresnel zone it belongs to, i.e.
\begin{equation}
	\phi_n  = \psi\left( \frac{l^\br_n+l^\ru_n}{2} \right)  .
	\label{phiback}
\end{equation}
 
The SP based algorithm is summarized in  {Algorithm \ref{alg1}}. In practice, the calculation of  $v(a)$ and $\psi(a)$ involves numerical integration over the interval  $[a_{\text{min}}, a_{\text{max}}]$, which is appropriately sampled. This partitioning proves effective if the spacing between two integral points is less than half of the RIS elements' spacing. Consequently, the number of samples on $a$ is proportional to the number of diagonal elements on the RIS, on the order of $\mathcal{O}(\sqrt{N^\text{RIS}})$. Therefore, throughout the steps of Algorithm \ref{alg1}, the computational complexity remains within $\mathcal{O}(N^\text{RIS})$.

\begin{algorithm}[htbp]
	
	\caption{\cb Near-field Wideband RIS Beamforming Based on Fresnel Zone and Stationary Phase Method (FZ-SPM)  }\label{alg:alg1}
	\cb
	\textbf{Inputs:} The location of BS $(x^\bs,y^\bs,z^\bs)$ and UE $(x^\ue,y^\ue,z^\ue)$; designed bandwidth $B$; center frequency $f_c$
	\\1. Translate the coordinate system by \eqref{rotation}.
	\\2.  Determine the semi-major axis $a$ of Fresnel zones with proper division in $[a_{\text{min}}, a_{\text{max}}]$, and calculate inner integral $v(a)$ by \eqref{va}.
	\\3. Obtain phase design across Fresnel zones $\psi(a)$ by \eqref{final_phase1}. 
	\\4. Calculate phase shift $\phi_n$ by \eqref{phiback} for $n=1,...,N^\ris$.
	\\ 	\textbf{Outputs:} The phase shift matrix $\bm{\Theta}=\text{diag}(\phi_1,...,\phi_n)$.
	\label{alg1}
\end{algorithm}
}

\subsection{\cb RIS Beamforming Based on Gerchberg-Saxton Algorithm}
{\cb
Although the SP method employs a nearly closed-form expression and acceptable performance, it only fits the condition that $B\cdot \Delta_t$ is sufficiently large \cite{fowle1964design}. In systems without huge size of RIS, which keeps $\Delta_t$ not significantly large, or with limited bandwidth $B$, the SP method may not generate desired gain spectrum, requiring an alternative method. 

To start with, the phase shift of different Fresnel zone is sampled as a vector $\mathbf{w} = [\mathrm{e}^{\mathrm{j}\psi_t(t_1)}, ... , \mathrm{e}^{\mathrm{j}\psi_t(t_{N_S})}]$, where $t_n = t_{\rm min} + \frac{n-1}{N_S-1}(t_{\rm max}-t_{\rm min})$ and $N_S$ is the number of samples. Similarly, the sampling on frequency is defined as $f_k = f_c-B'/2+kB'/K', k = 1, ..., K'$, where $K'$ is the number of samples and $B'>B$ is the extended bandwidth which is used for minimizing the out-of-band leakage. The sampled gain vector $\mathbf{g}$ is defined as $[\mathbf{g}]_k = g(f_k)$. Then the relationship in \eqref{ftransform} can be expressed in discrete domain as
\begin{equation}
	\begin{aligned}
		\mathbf{g} = \mathbf{Aw}, 
	\end{aligned}
\end{equation} 
where $[\mathbf{A}]_{(k,n)} = \mathrm{e}^{-\mathrm{j}2\pi f_k t_n}v_t(t_n) $. The aim our beamforming design is make the gain spectrum $\mathbf{g}$ as close as the ideal gain spectrum $\hat{\mathbf{g}}$. The objective of the phase shift vector $\mathbf{w}$ design can be expressed as
\begin{equation}
	\begin{aligned}
		\min_{\mathbf{w}} & \|\hat{\mathbf{g}}-\mathbf{g}\|_2 \\
		s.t. & |[]\mathbf{w}]_n| = 1, n = 1,..., N_S,
	\end{aligned}
	\tag{P1}
	\label{objective}
\end{equation}
where $[\hat{\mathbf{g}}]_k = g(f_k)$ and $g(f)$ is described in \eqref{ideal}.  

The problem in \eqref{objective} can be regarded as a phase retrieval problem \cite{PhaseRetrieval2015}, which can be solved by Gerchberg-Saxton (GS) algorithm \cite{lu2023hierarchical}. The GS algorithm is based on alternating projections. The beamforming design based on the GS algorithm is summarized in {Algorithm \ref{alg2}}. 

In the GS algorithm, we use ${\mathbf{g}'}_{(m)}$, ${\mathbf{g}}_{(m)}$, $\mathbf{w}_{(m)}$, and $\mathbf{w}'_{(m)}$ to denote the gain with desired value, the gain generated by designed phase shifts, the designed phase shift vector and the phase shift vector obtained by revised gain in the $m$-th iteration of the GS algorithm.

The initialization of the vector $\mathbf{w}_{(0)}$ is crucial to the overall performance because the solution to the non-convex problem \eqref{objective} is highly sensitive to the choice of the initial value. Typically, the initial phase of $\mathbf{w}$ is generated randomly, which often leads to convergence at local minima, though it occasionally reaches the global minimum. An alternative approach is to utilize an existing beamforming vector, such as one derived from the SP method. This method generally guarantees a certain level of performance, but it may still fall short of achieving the global minimum. In this study, we opt for the latter approach, since the performance achieved by the SP method is already near the upper bound, as demonstrated later in the paper.

In $m$-th iteration, with given $\mathbf{w}_{(m)}$, the beamforming gain is written as
\begin{equation}
	\mathbf{g}_{(m)} = \mathbf{Aw}_{(m)}.
\end{equation}
Here, $\mathbf{g}_{(m)}$ do not have the value of ideal gain. What to do in the next step is to keep the phase of the designed gain, and change the amplitude to the ideal one.  In this case, $k$-th entry of $\mathbf{g}'_{(m)}$ can be expressed as
\begin{equation}
	[\mathbf{g}'_{(m)}]_k  = \frac{[\mathbf{g}_{(m)}]_k}{\left|[\mathbf{g}_{(m)}]_k\right|} [\hat{\mathbf{g}}_{(m)}]_k.
\end{equation}
The third step is to find an unconstrained $\mathbf{w}'_{(m)}$ to minimize $\|\mathbf{g}'_{(m)}-\mathbf{Aw}'_{(m)}\|_2$, which can be obtained by least square (LS) algorithm as
\begin{equation}
	\mathbf{w}'_{(m)} = (\mathbf{A}^H\mathbf{A})^{-1}\mathbf{A}^H\mathbf{g}'_{(m)}. 
\end{equation}
 Finally, we make $\mathbf{w}'_{(m+1)}$ to follow the unit-modulus constraint, with $n$-th entry represented as
 \begin{equation}
 	[\mathbf{w}_{(m+1)}]_n = [\mathbf{w}'_{(m)}]_n/\left|[\mathbf{w}'_{(m)}]_n \right|
 \end{equation}
 After the number of iterations reaches $m_{\rm max}$, we utilize $\mathbf{w}_{(m_{\rm max})}$ as the designed beamforming vector. The designed phase at sample points can therefore expressed as $\psi_t^{GS}(t_n) = \angle [\mathbf{w}_{(m_{\rm max})}]_n$. Using interpolation we can get the designed phase on different Fresnel zones. Finally, the phase design of RIS elements can be expressed as
 \begin{equation}
 	\phi^{GS}_n  = \psi_t^{GS}\left( \frac{l^\br_n+l^\ru_n}{c} \right).
 \end{equation}
 
Although the GS algorithm exhibits much higher computational complexity compared with the SP method, it remains manageable since the algorithm operates solely on one dimension across different Fresnel zones rather than the entire RIS plane. Specifically, when considering the number of iterations as a constant, the complexity of the GS algorithm is $\mathcal{O}(N_S^3+K'N_S^2)$. Given that the number of samples $N_S$ is on the order of $\sqrt{N^\ris}$, the complexity of the GS algorithm can be further characterized as $\mathcal{O}\left(({N^\ris}\right)^{1.5}+K'{N^\ris})$. 
 

}

\begin{algorithm}[htbp]
	
	\caption{\cb Near-field Wideband RIS Beamforming Based on Fresnel Zones and GS Algorithm  (FZ-GSA)}\label{alg:alg1}
	{\cb
	\textbf{Inputs:} $\mathbf{A}$, $\mathbf{w}_{(0)}$, $\hat{\mathbf{g}}$
	\\For $m = 1,..., m_{\rm max}$ 	
	\\ ${~~~~}$ $\mathbf{g}_{(m)} = \mathbf{Aw}_{(m)}$,
	\\${~~~~}$ $[\mathbf{g}'_{(m)}]_k  = \frac{[\mathbf{g}_{(m)}]_k}{\left|[\mathbf{g}_{(m)}]_k\right|} [\hat{\mathbf{g}}_{(m)}]_k$,
	\\${~~~~}$ $\mathbf{w}'_{(m)} = (\mathbf{A}^H\mathbf{A})^{-1}\mathbf{A}^H\mathbf{g}'_{(m)}$,
	\\${~~~~}$ $[\mathbf{w}_{(m+1)}]_n = [\mathbf{w}'_{(m)}]_n/\left|[\mathbf{w}'_{(m)}]_n \right|$,
	\\end
	\\$\psi_t^{GS}(t_n) = \angle \left[\mathbf{w}_{(m_{\rm max})}\right]_n$
	\\$	\phi^{GS}_n  = \psi_t^{GS}\left( \frac{l^\br_n+l^\ru_n}{c} \right)$
	\\ 	\textbf{Outputs:} The phase shift matrix $\bm{\Theta^{GS}}=\text{diag}(\phi^{GS}_1,...,\phi^{GS}_n)$.
}
	\label{alg2}
\end{algorithm}

\section{Simulation Results}{\label{result}}
In this section, simulation results are provided to validate the effectiveness of the proposed wideband beamforming methods based on Fresnel zone.

\subsection{Simulation Setup}

We consider a near-field RIS-aided system with a square planar RIS with side length $D=1$ m. The carrier frequency $f_c$ is set as 30 GHz. We use half-wavelength spacing of the RIS elements \cite{hao_ultra_2022,su_wideband_2023}. The bandwidth is set to $B = 1.5$ GHz. We assume that the power spectral density of noise is -170 dBm/Hz \cite{myers_infocus_2022}. It is assumed that the location of the TX and the RX, equivalently the channel $h^\br(x,y,f)$ and $h^\ru(x,y,f)$ are known at the RIS controller, which can be acquired by the algorithms in \cite{lu_near-field_2023,wei_codebook_2022,yang_channel_2023,pan_ris-aided_2023}.

\begin{figure}[t]
	\centering
	\includegraphics[width=0.5\textwidth]{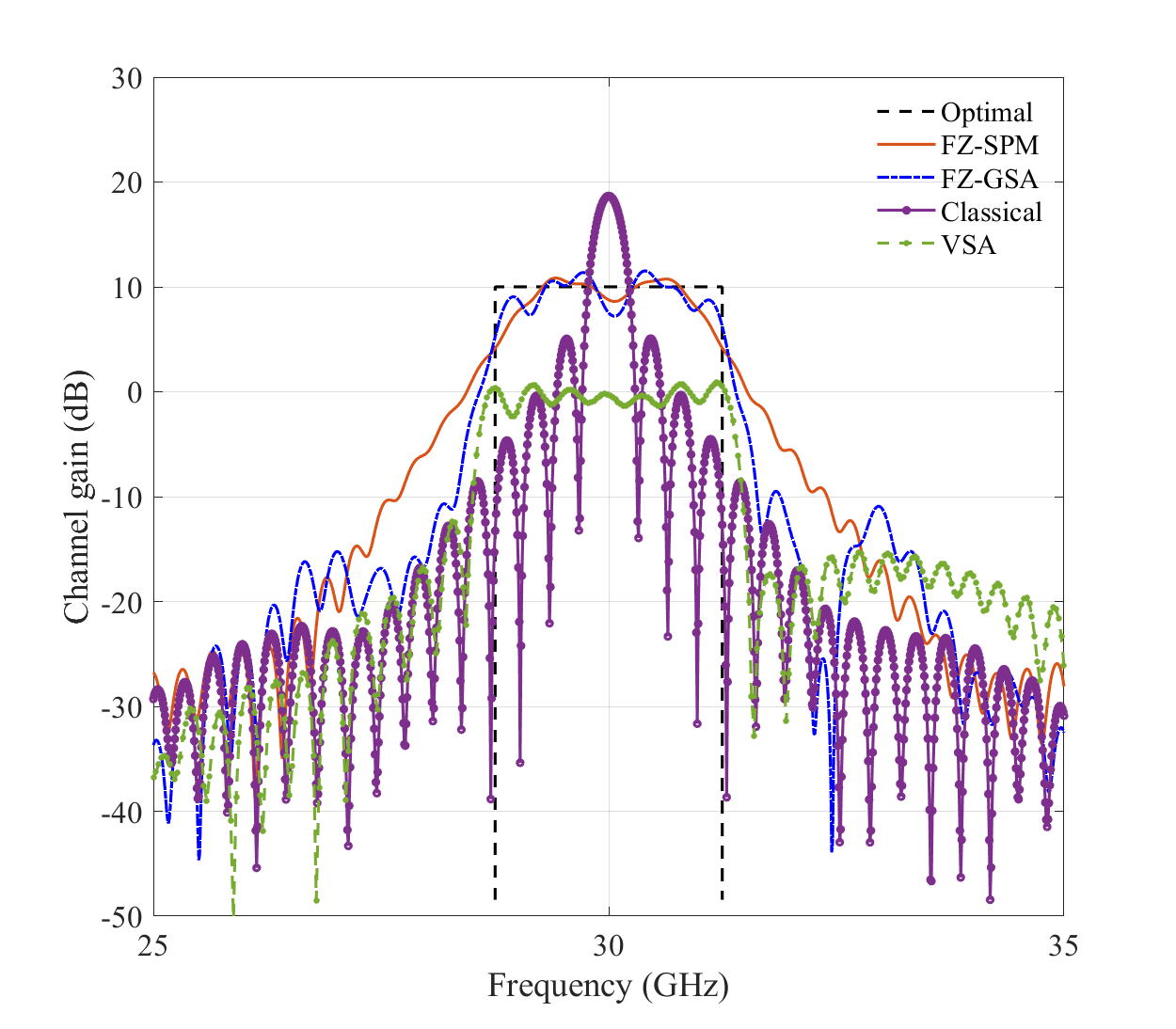}
	\caption{Beamforming gain at different frequencies}
	\label{fig:beamforming_performance}
\end{figure}

\subsection{Beamforming Performance}

The beamforming performance of our Frsenel zone-based wideband beamforming is illustrated in this subsection.

We set the TX at the location $(x^\bs, y^\bs, z^\bs) = (6.4~\mathrm{m}, 5 ~\mathrm{m} ,14.4 ~\mathrm{m} )$, and the RX at the location $(x^\ue, y^\ue, z^\ue) = (-4.8~ \mathrm{m}, 5 ~\mathrm{m} ,6.4 ~\mathrm{m} )$ .  The beamforming gain at different frequencies is shown in Fig. \ref{fig:beamforming_performance}. The optimal beamforming distributes energy evenly across the entire frequency band as described in \eqref{ideal}, which is hardly achieved due to the modular constraints of RIS elements. {\cb It can be observed that the beamforming gain of both the proposed FZ-SPM and FZ-GSA  approaches based on the Fresnel zone model is approximately flat over the frequency band, close to the optimal beamforming. The FZ-GSA based beamforming has flatter in-band gain and lower out-of-band leakage, while the FZ-SPM based beamforming enjoys lower computational complexity. Compared with the classical narrowband beamforming, the proposed methods can mitigate the loss in the edge frequency and overcome the near-field beam split effect.} Although the VSA based beamforming can also form flat in-band gain, it suffers server loss in gain due to its separate design of each sub-array.  

\subsection{Achievable Rate Performance}
In this subsection, the performance on average achievable rate based on the proposed method is provided with the consideration on the near-field beam split effect.

\begin{figure}[t]
	\centering
	\includegraphics[width=0.5\textwidth]{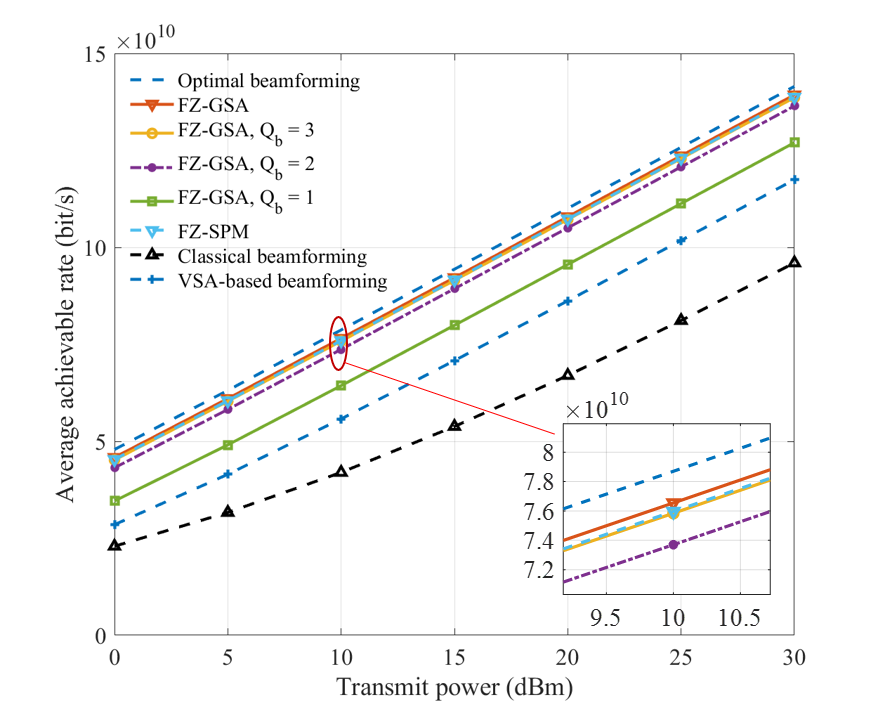}
	\caption{Simulation results of average achievable rate vs. transmit power}
	\label{fig:sum_rate_vs_power}
\end{figure}

In each random experiment, the TX and the RX are randomly located in the distance range $l^\br\in \left[ l^\br_{\text{min}}, l^\br_{\text{max}}\right]$ and $l^\ru\in \left[ l^\ru_{\text{min}}, l^\ru_{\text{max}}\right]$, respectively. We set $l^\br_{\text{min}}=l^\ru_{\text{min}} = 7 \mathrm{m}$ and $l^\br_{\text{max}}=l^\ru_{\text{max}} = 13 \mathrm{m}$ unless otherwise mentioned. We first investigate the average achievable rate as a function of transmit power, as shown in Fig. \ref{fig:sum_rate_vs_power}. The optimum beamforming gain described in \eqref{ideal} represents the upper limit, against which the standard narrowband beamforming and the VSA-based method in \cite{cheng_achievable_2024} serves as the baseline. {\cb The proposed FZ-SPM and FZ-GSA methods exhibit around 50\% increase in achievable rate compared to the classical beamforming and a 30\% increase compared to the VSA-based beamforming, while maintaining a marginal deficit of less than 5\% relative to the optimal benchmark. The performance gap between FZ-SPM and FZ-GSA is minimal, less than 1\%. Therefore, choosing FZ-SPM for real-world applications seems more practical because it has lower computational complexity. Furthermore, we analyze the performance using discrete resolution phase shifts with resolution values $Q_b$ of 1, 2, and 3 bits. The findings indicate that with only 2-bit phase shifts, the proposed methods nearly match the performance of continuous phase shifts, while the former offers advantages in terms of lower hardware complexity and cost.}

\begin{figure}[t]
	\centering
	\includegraphics[width=0.45\textwidth]{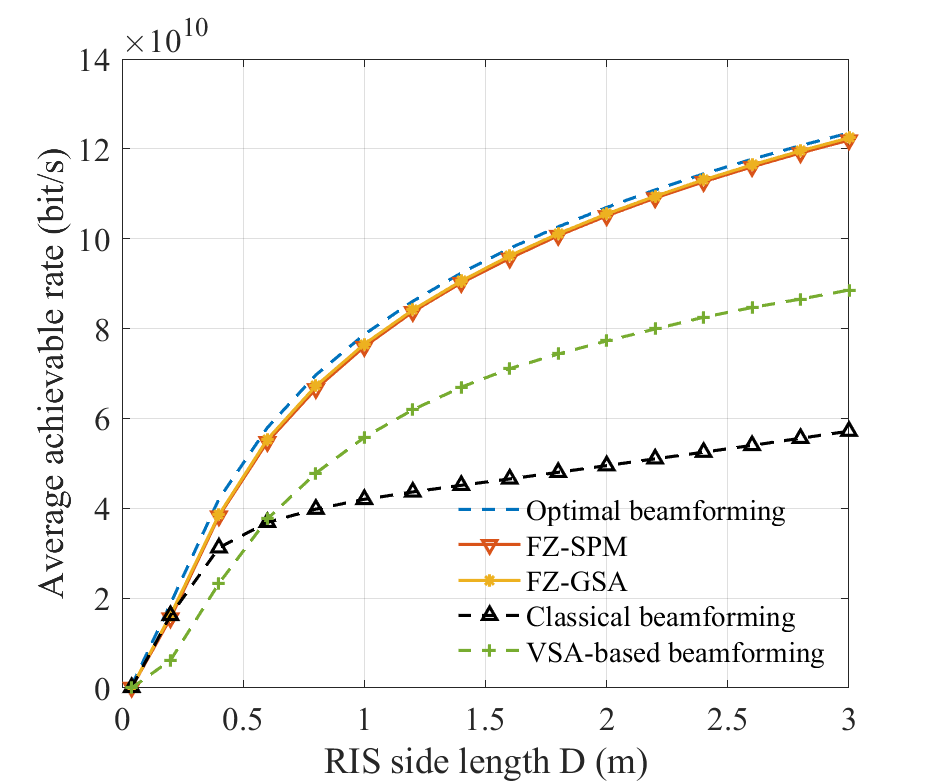}
	\caption{Simulation results of average achievable rate vs. side length of RIS}
	\label{fig:sum_rate_vs_D}
\end{figure}

\begin{figure}[t]
	\centering
	\includegraphics[width=0.45\textwidth]{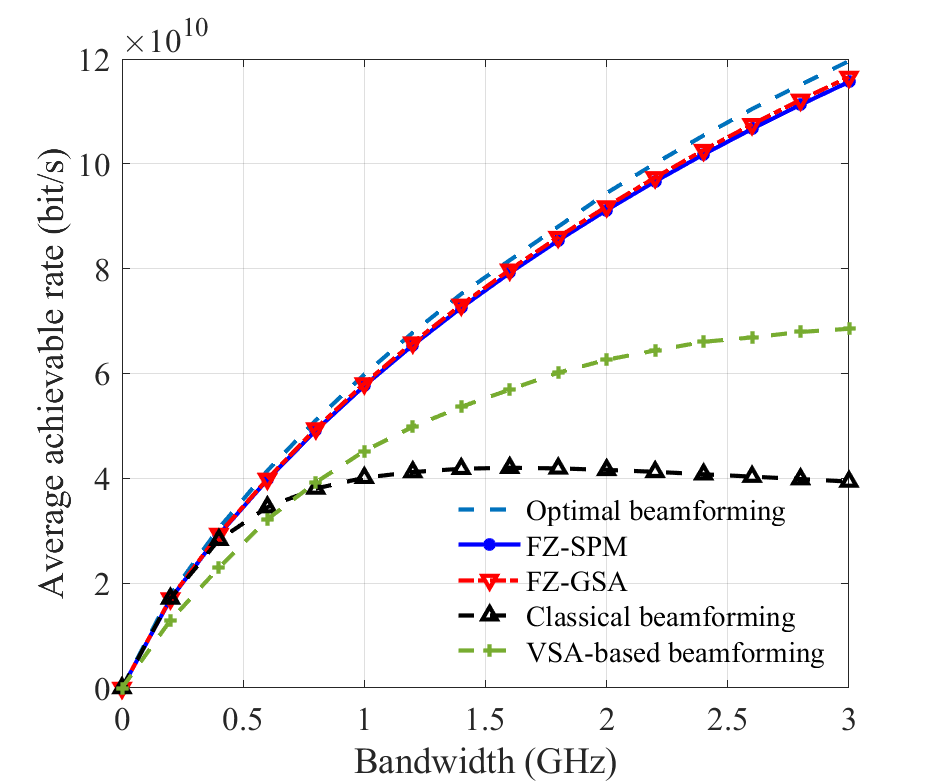}
	\caption{Simulation results of average achievable rate vs. system bandwidth}
	\label{fig:sum_rate_vs_bandwidth}
\end{figure}

Additionally, we present the average achievable rate as a function of the size of the RIS and the bandwidth in Fig. \ref{fig:sum_rate_vs_D} and Fig. \ref{fig:sum_rate_vs_bandwidth}, respectively. The results reveal that the achievable rate of classical beamforming experiences minimal improvement with increasing RIS size and eventually reaches a plateau as the bandwidth widens. By employing proposed wideband beamforming design, {\cb our proposed methods} effectively mirror the advancement of optimal beamforming. Consequently, our proposed beamforming techniques facilitate the full exploitation of the performance benefits of wideband RIS, without succumbing to the near-field wideband beam split effect.

\begin{figure}[t]
	\centering
	\includegraphics[width=0.45\textwidth]{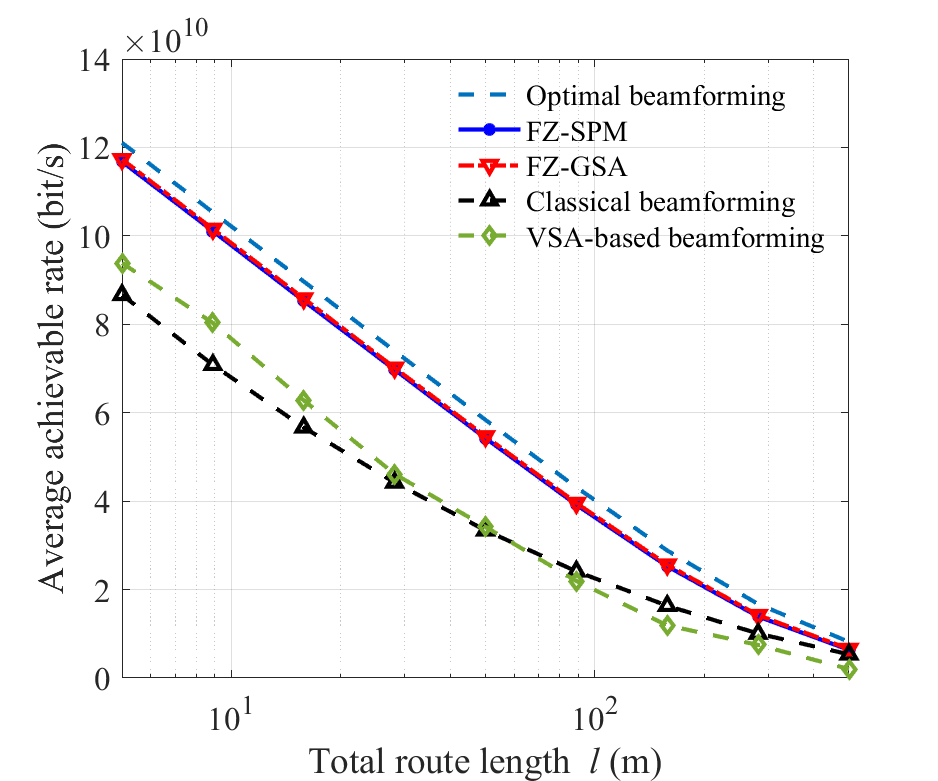}
	\caption{Simulation results of average achievable rate vs. transmit distance}
	\label{fig:sum_rate_vs_distance}
\end{figure}

Moreover, we highlight the influence of the total communication route length on the average achievable rate in Fig. \ref{fig:sum_rate_vs_distance}.  Results demonstrate that our proposed beamforming approaches retain their advantages across both near-field and far-field scenarios, yielding a 60\% higher rate compared to classical beamforming at a route length of as large as 200 m. Given that the far-field channel model is a specialized example of the near-field channel model \cite{cui2022channel}, it is unsurprising that our near-field wideband beamforming solution performs well in the far-field context. 

\begin{figure}[t]
	\centering
	\includegraphics[width=0.45\textwidth]{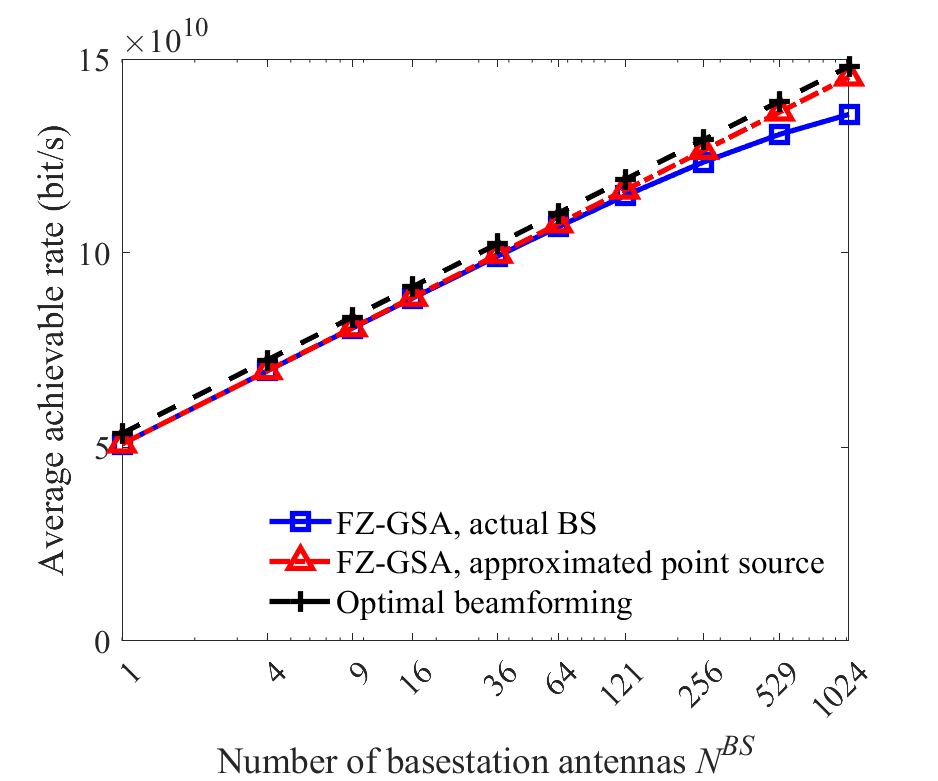}
	\caption{Simulation results of average achievable rate vs. number of BS antennas}
	\label{fig:sum_rate_vs_nBS}
\end{figure}

{\cb Finally, we conduct simulations to assess the average achievable rate considering BS with multiple antennas. In Section II-A, we assume that the RIS is positioned in the far field of the BS. This assumption allows us to treat the BS as if it has antennas concentrated at a single point. We compare the performance of our proposed FZ-GSA beamforming in the actual BS model and the approximated model.  Our analysis, depicted in Fig. \ref{fig:sum_rate_vs_nBS}, reveals a strong alignment between the approximation and the actual scenario when the number of antennas $N^\bs$ is below 500, with a marginal performance degradation of less than 4\%. This outcome underscores the robustness of our beamforming design considering multiple BS antennas.}

\section{Conclusion}{\label{conclusion}}
In this work, we have addressed the near-field beam split effect in RIS-aided wideband communications, a critical issue that leads to significant beamforming gain loss. To address this problem, we have proposed a novel near-field wideband RIS beamforming method based on Fresnel zones, marking the first introduction of the principle of Fresnel zones from electromagnetic measurement into RIS communications. This approach offers a fresh perspective for RIS beamforming, leveraging the inherent properties of Fresnel zones to enhance the performance of wireless communication systems. Through simulations, we have demonstrated that our methods can effectively mitigate the near-field beam split effect, resulting in a uniform gain across the entire frequency band. This is achieved without increasing the hardware complexity or cost, making it suitable for practical implementation in RIS systems. 

In addressing the beamforming problem to overcome the near-field beam splitting effect in RIS, this paper has focused solely on the optimization objective of maximizing the achievable rate. Other significant performance metrics, such as energy efficiency\cite{you2021energy}, total transmit power \cite{zhang2023active}, and physical layer security rate \cite{zhang2024physical}, could be introduced as evaluation criteria for future research on near-field wideband beamforming design. Moreover, this study only considered the simple scenario of single-user communication. Further research could extend to more complex scenarios involving multiple-antenna, multi-user, and cell-free network scenarios \cite{zhang2021joint}, refining the beamforming designs to accommodate these advanced configurations. 

\bibliographystyle{IEEEtran}
\bibliography{IEEEabrv,RIS_beamforming}

\vfill

\end{document}